\documentclass[10pt,a4paper,onecolumn]{IEEEtran}
\usepackage[lmargin=2cm, rmargin=2cm, tmargin=2cm, bmargin=2cm]{geometry}
\usepackage{ccaption}
\usepackage[T1]{fontenc}
\usepackage{graphicx}
\usepackage{fancyhdr}
\usepackage{helvet}
\usepackage{floatflt}
\usepackage{colortbl}
\usepackage{fancyvrb}
\usepackage{wrapfig}
\usepackage[round,authoryear]{natbib}
\usepackage[latin1]{inputenc}
\usepackage[T1]{fontenc}
\usepackage[english]{babel}
\usepackage{graphicx}
\usepackage{subfigure}
\usepackage{amsfonts}
\usepackage{amsmath}
\usepackage{theorem}
\usepackage{ae}

\newtheorem{assumption}{Assumption}
\newtheorem{remark}{Remark}
\newtheorem{theorem}{Theorem}

\DeclareMathOperator{\sgn}{sgn}
\DeclareMathOperator{\sat}{sat}

\title{An adaptive fuzzy sliding mode controller for nonlinear systems with non-symmetric dead-zone and its application to an electro-hydraulic system}
\author{Wallace Moreira Bessa}
\date{}

\begin{document}

\maketitle

\abstract{The dead-zone is one of the most common hard nonlinearities in industrial 
actuators and its presence may drastically compromise control systems stability and performance. In 
this work, an adaptive variable structure controller is proposed to deal with a class of uncertain 
nonlinear systems subject to a non-symmetric dead-zone input. The adopted approach is primarily 
based on the sliding mode control methodology but enhanced by an adaptive fuzzy algorithm to 
compensate the dead-zone. Using Lyapunov stability theory and Barbalat's lemma, the convergence 
properties of the closed-loop system are analytically proven. In order to illustrate the controller 
design methodology, an application of the proposed scheme to an electro-hydraulic system is introduced. 
The performance of the control system is evaluated by means of numerical simulations.}

\section{INTRODUCTION}

Dead-zone is a hard nonlinearity, frequently encountered in many actuators of industrial control 
systems, especially those containing some very common components, such as hydraulic \cite{knohl1,%
conem2006,valdiero1} or pneumatic \cite{guenther1,andrighetto1,valdiero2} valves. Dead-zone 
characteristics are often unknown and it was already observed that its presence can severely 
reduce control system performance and lead to limit cycles in the closed-loop system. 

The growing number of papers involving systems with dead-zone input confirms the importance of 
taking such a non-smooth nonlinearity into account during the control system design process. 
The most common approaches are adaptive schemes \cite{tao1,wang1,zhou1,ibrir1}, fuzzy systems 
\cite{kim1,oh1,lewis1,conem2008}, neural networks \cite{selmic1,tsai1,zhang1} and variable 
structure methods \cite{corradini1,shyu1}. Many of these works \cite{tao1,kim1,oh1,selmic1,%
tsai1,zhou1} use an inverse dead-zone to compensate the negative effects of the dead-zone 
nonlinearity even though this approach leads to a discontinuous control law and requires 
instantaneous switching, which in practice can not be accomplished with mechanical actuators. 
An alternative scheme, without using the dead-zone inverse, was originally proposed by 
\citep{lewis1} and also adopted by \citep{wang1}. In both works, the dead-zone is 
treated as a combination of a linear and a saturation function. This approach was further extended 
by \citep{ibrir1} and by \citep{zhang1}, in order to accommodate non-symmetric and unknown 
dead-zones, respectively.

On this basis, sliding mode control can be considered as a very attractive approach because of its 
robustness against both structured and unstructured uncertainties as well as external disturbances. 
Nevertheless, the discontinuities in the control law must be smoothed out to avoid the undesirable 
chattering effects. The adoption of properly designed boundary layers have proven effective in 
completely eliminating chattering, however, leading to an inferior tracking performance. 

As demonstrated by \citep{tese,rsba2010,nd2012,Bessa2019}, adaptive fuzzy algorithms can be properly embedded in 
smooth sliding mode controllers to compensate for modeling inaccuracies, in order to improve the 
trajectory tracking of uncertain nonlinear systems. It has also been shown that adaptive fuzzy 
sliding mode controllers are suitable for a variety of applications ranging from underwater 
robotic vehicles \cite{diname2007,raas2008} to the chaos control in a nonlinear pendulum 
\cite{cba2008b,csf2009}.

As a matter of fact, intelligent control has proven to be a very attractive approach to cope with uncertain nonlinear systems 
\citep{cobem2005,Bessa2017,Bessa2018,Lima2018,Deodato2019,Lima2020}. 
By combining nonlinear control techniques, such as feedback linearization or sliding modes, with adaptive intelligent algorithms, 
for example fuzzy logic or artificial neural networks, the resulting intelligent control strategies can deal with the nonlinear 
characteristics as well as with modeling imprecisions and external disturbances that can arise.

In this paper, an adaptive fuzzy sliding mode controller is proposed to deal with uncertain nonlinear
systems subject to a non-symmetric dead-zone input. The adopted control scheme is primarily based on 
the sliding mode control methodology, but an adaptive fuzzy inference system is introduced to 
compensate for dead-zone effects. Based on a Lyapunov-like analysis using Barbalat's lemma, the 
convergence properties of the closed-loop signals are analytically proven. An application of the 
proposed control strategy to a third order nonlinear system (electro-hydraulic system) is introduced 
to illustrate the controller design process. Simulation studies are also presented in order to 
demonstrate the control system performance.

\section{PROBLEM STATEMENT}

\begin{figure}[hbt]
\centering
\includegraphics[width=0.4\textwidth]{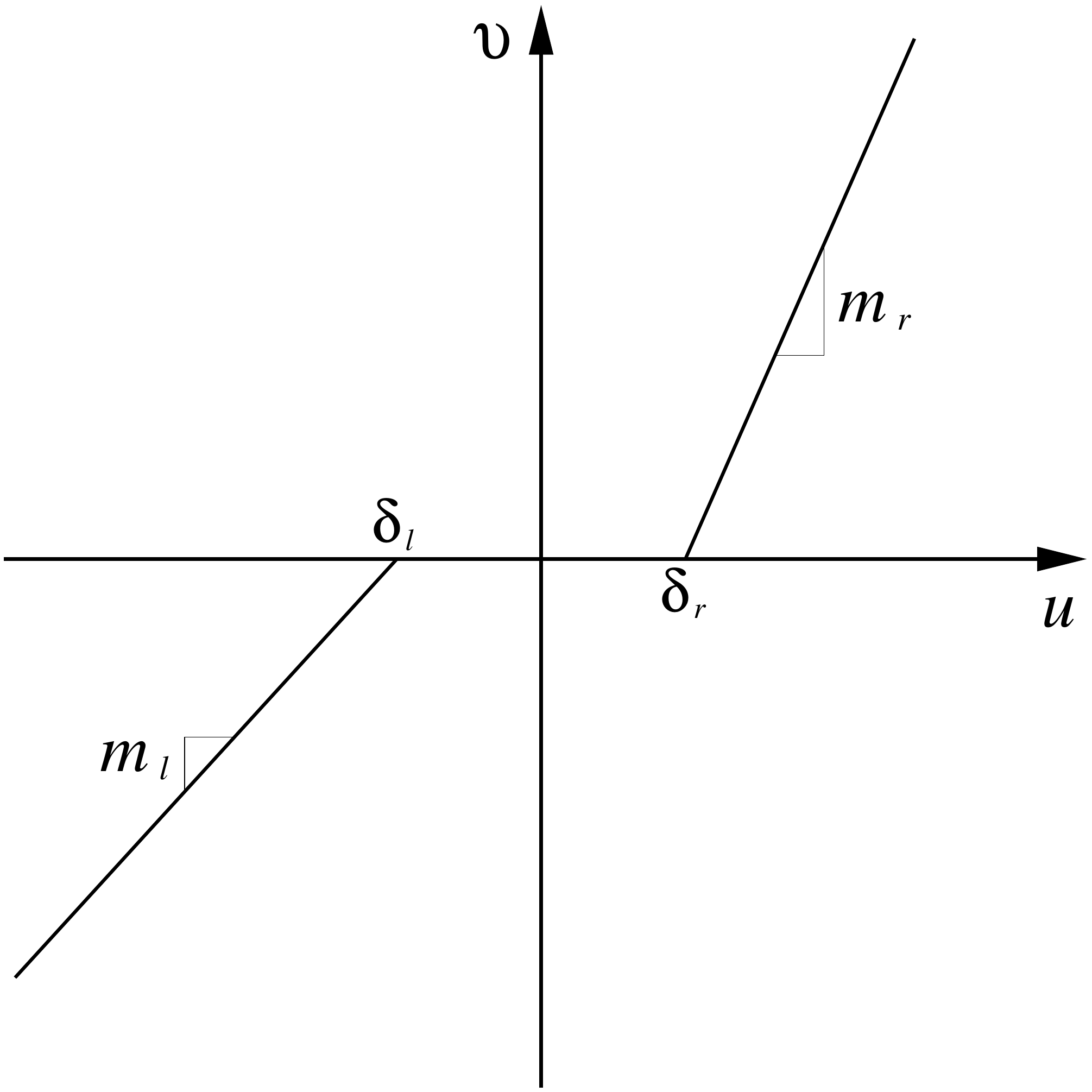} 
\caption{Dead-zone nonlinearity.}
\label{fi:dzone}
\end{figure}

Consider a class of $n^\mathrm{th}$-order nonlinear system:

\begin{equation}
x^{(n)}=f(\mathbf{x})+b(\mathbf{x})\upsilon
\label{eq:system}
\end{equation}

\noindent
where the scalar variable $x\in\mathbb{R}$ is the output of interest, $x^{(n)}\in\mathbb{R}$ is the 
$n^\mathrm{th}$ derivative of $x$ with respect to time $t\in[0,+\infty)$, $\mathbf{x}=[x,\dot{x},
\ldots,x^{(n-1)}]\in\mathbb{R}^n$ is the system state vector, $f,b:\mathbb{R}^n\to\mathbb{R}$ are 
both nonlinear functions and $\upsilon\in\mathbb{R}$ represents the output of a dead-zone function 
$\Upsilon:\mathbb{R}\to\mathbb{R}$, as shown in Fig.~\ref{fi:dzone}, with $u\in\mathbb{R}$ stating
for the controller output variable.

The adopted dead-zone model is mainly based on that proposed in \cite{ibrir1}, which can be 
mathematically described by 

\begin{equation}
\upsilon=\Upsilon(u)= \left\{\begin{array}{ll}
m_l(u-\delta_l)&\mbox{if}\quad u\le\delta_l\\
0&\mbox{if}\quad  \delta_l<u<\delta_r\\
m_r(u-\delta_r)&\mbox{if}\quad u\ge\delta_r 
\end{array}\right.
\label{eq:dzone1}
\end{equation}

In respect of the dead-zone model presented in Eq.~(\ref{eq:dzone1}), the following assumptions can 
be made: 

\begin{assumption}
The dead-zone output $\upsilon$ is not available to be measured.
\label{as:output}
\end{assumption}
\begin{assumption}
The dead-band parameters $\delta_l$ and $\delta_r$ are unknown but bounded and with known signs, 
i.e., $\delta_{l\,\mathrm{min}}\le\delta_l\le\delta_{l\,\mathrm{max}}<0$ and $0<\delta_{r\,
\mathrm{min}}\le\delta_r\le\delta_{r\,\mathrm{max}}$.
\label{as:dband}
\end{assumption}
\begin{assumption}
The slopes in both sides of the dead-zone are unknown but positive and bounded, i.e., $0<m_{l\,
\mathrm{min}}\le m_l\le m_{l\,\mathrm{max}}$ and $0<m_{r\,\mathrm{min}}\le m_r\le m_{r\,
\mathrm{max}}$
\label{as:slopes}
\end{assumption}

For control purposes, Eq.~(\ref{eq:dzone1}) can be rewritten in a more appropriate form:

\begin{equation}
\upsilon=\Upsilon(u)=m(u)[u-d(u)] 
\label{eq:dzone2}
\end{equation}

\noindent
where 

\begin{equation}
m(u)= \left\{\begin{array}{ll}
m_l&\mbox{if}\quad u\le0\\
m_r&\mbox{if}\quad u>0 
\end{array}\right.
\label{eq:slopes}
\end{equation}

\noindent
and

\begin{equation}
d(u)= \left\{\begin{array}{ll}
\delta_l&\mbox{if}\quad u\le\delta_l\\
u&\mbox{if}\quad  \delta_l<u<\delta_r\\
\delta_r&\mbox{if}\quad u\ge\delta_r 
\end{array}\right.
\label{eq:dsat}
\end{equation}

\begin{remark}
From Assumption~\ref{as:dband} and Eq.~(\ref{eq:dsat}), it can be easily verified that $d(u)$ 
is bounded: $|d(u)|\le\delta$, where $\delta=\mathrm{max}\{-\delta_{l\,\mathrm{min}},\delta_{r\,
\mathrm{max}}\}$.
\end{remark}

In respect of the dynamic system presented in Eq.~(\ref{eq:system}), the following assumptions can 
also be made:

\begin{assumption}
The function $f$ is unknown but bounded by a known function of $\mathbf{x}$, i.e., $|\hat{f}
(\mathbf{x})-f(\mathbf{x})|\le F(\mathbf{x})$ where $\hat{f}$ is an estimate of $f$.
\label{as:fbounds}
\end{assumption}
\begin{assumption}
The input gain $b(\mathbf{x})$ is unknown but positive and bounded, i.e., $0<b_{\mathrm{min}}
\le b(\mathbf{x})\le b_{\mathrm{max}}$.
\label{as:bbounds}
\end{assumption}

\section{CONTROLLER DESIGN}

The proposed control problem is to ensure that, even in the presence of parametric uncertainties, 
unmodeled dynamics and a non-symmetric dead-zone input, the state vector $\mathbf{x}$ will follow a 
desired trajectory $\mathbf{x}_d=[x_d,\dot{x}_d,\ldots,x^{(n-1)}_d]$ in the state space.

Regarding the development of the control law, the following assumptions should also be made:

\begin{assumption}
The state vector $\mathbf{x}$ is available.
\label{as:stat}
\end{assumption}
\begin{assumption}
The desired trajectory $\mathbf{x}_d$ is once differentiable in time. Furthermore, every element
of vector $\mathbf{x}_d$, as well as $x^{(n)}_d$, is available and with known bounds.
\label{as:traj}
\end{assumption}

Now, let $\tilde{x}=x-x_d$ be defined as the tracking error in the variable $x$, and 

\begin{displaymath}
\mathbf{\tilde{x}}=\mathbf{x}-\mathbf{x}_d=[\tilde{x},\dot{\tilde{x}},\ldots,\tilde{x}^{(n-1)}]
\end{displaymath}

\noindent 
as the tracking error vector. 

Consider a sliding surface $S$ defined in the state space by the equation $s(\mathbf{\tilde{x}})=0$, 
with the function $s:\mathbb{R}^n\rightarrow\mathbb{R}$ satisfying

\begin{displaymath}
\displaystyle s(\mathbf{\tilde{x}})=\left(\frac{d}{dt}+\lambda\right)^{n-1}\tilde{x}
\end{displaymath}

\noindent 
or conveniently rewritten as

\begin{equation}
s(\mathbf{\tilde{x}})=\mathbf{c^\mathrm{T}\tilde{x}}
\label{eq:surf}
\end{equation}

\noindent
where $\mathbf{c}=[c_{n-1}\lambda^{n-1},\ldots,c_1\lambda,c_0]$ and $c_i$ states for binomial 
coefficients, i.e.,

\begin{equation}
c_i=\binom{n-1}{i}=\frac{(n-1)!}{(n-i-1)!\:i!}\:,\quad i=0,1,\ldots,n-1 
\label{eq:binom}
\end{equation}

\noindent
which makes $c_{n-1}\lambda^{n-1}+\cdots+c_1\lambda+c_0$ a Hurwitz polynomial. 

From Eq.~(\ref{eq:binom}), it can be easily verified that $c_0=1$, for $\forall n\ge1$. Thus, for 
notational convenience, the time derivative of $s$ will be written in the following form:

\begin{equation}
\dot{s}=\mathbf{c^\mathrm{T}\dot{\tilde{x}}}
=\tilde{x}^{(n)}+\mathbf{\bar{c}^\mathrm{T}\tilde{x}}
\label{eq:sd}
\end{equation}

\noindent
where $\mathbf{\bar{c}}=[0,c_{n-1}\lambda^{n-1},\ldots,c_1\lambda]$.

Now, let the problem of controlling the uncertain nonlinear system (\ref{eq:system}) be treated in
a Filippov's way \cite{filippov1}, defining a control law composed by an equivalent control $\hat{u}
=\widehat{bm}^{-1}(-\hat{f}+x^{(n)}_d-\mathbf{\bar{c}^\mathrm{T}\tilde{x}})$, an estimate $\hat{d}
(\hat{u})$ and a discontinuous term $-K\sgn(s)$:

\begin{equation}
u=\widehat{bm}^{-1}(-\hat{f}+x^{(n)}_d-\mathbf{\bar{c}^\mathrm{T}\tilde{x}})
+\hat{d}(\hat{u})-K\sgn(s)
\label{eq:usgn}
\end{equation}

\noindent
where $\widehat{bm}=\sqrt{b_\mathrm{max}m_\mathrm{max}b_\mathrm{min}m_\mathrm{min}}$ with 
$m_\mathrm{max}=\mathrm{max}\{m_{l\,\mathrm{max}},m_{r\,\mathrm{max}}\}$ and $m_\mathrm{min}=
\mathrm{min}\{m_{l\,\mathrm{min}},m_{r\,\mathrm{min}}\}$, $K$ is a positive gain and $\sgn(\cdot)$ 
is defined as 

\begin{displaymath}
\sgn(s) = \left\{\begin{array}{rc}
-1&\mbox{if}\quad s<0 \\
0&\mbox{if}\quad s=0 \\
1&\mbox{if}\quad s>0
\end{array}\right.
\end{displaymath}

Based on Assumptions~\ref{as:dband}--\ref{as:bbounds} and considering that $\beta^{-1}\le
\widehat{bm}/(bm)\le\beta$, where $\beta=\sqrt{(b_\mathrm{max}m_\mathrm{max})/(b_\mathrm{min}
m_\mathrm{min})}$, the gain $K$ should be chosen according to

\begin{equation}
K\ge\beta\widehat{bm}^{-1}(\eta+F)+\delta+|\hat{d}(\hat{u})|+(\beta-1)|\hat{u}|
\label{eq:gain}
\end{equation}

\noindent
where $\eta$ is a strictly positive constant related to the reaching time. 

Therefore, it can be easily verified that (\ref{eq:usgn}) is sufficient to impose the sliding 
condition 

\begin{displaymath}
\displaystyle\frac{1}{2}\frac{d}{dt}s^2\le-\eta|s|
\end{displaymath}

\noindent
which, in fact, ensures the finite-time convergence of the tracking error vector to the sliding 
surface $S$ and, consequently, its exponential stability.

In order to obtain a good approximation to $d(u)$, the estimate $\hat{d}(\hat{u})$ will be computed
directly by an adaptive fuzzy algorithm.

The adopted fuzzy inference system is the zero order TSK (Takagi--Sugeno--Kang), whose rules can be 
stated in a linguistic manner as follows \cite{jang1}:

\begin{center}
\textit{If $\hat{u}$ is $\hat{U}_r$ then} $\hat{d}=\hat{D}_r\:,\quad r=1,2,\ldots,N$ 
\end{center}

\noindent
where $\hat{U}_r$ are fuzzy sets, whose membership functions could be properly chosen, and 
$\hat{D}_r$ is the output value of each one of the $N$ fuzzy rules.

Considering that each rule defines a numerical value as output $\hat{D}_r$, the final output 
$\hat{d}$ can be computed by a weighted average: 

\begin{equation}
\hat{d}(\hat{u}) = \frac{\sum_{r=1}^{N} w_r \cdot \: \hat{d}_r}{\sum_{r=1}^{N} w_r}
\label{eq:dcmean}
\end{equation}

\noindent
or, similarly,

\begin{equation}
\hat{d}(\hat{u}) = \mathbf{\hat{D}}^{\mathrm{T}}\mathbf{\Psi}(\hat{u})
\label{eq:dcvector}
\end{equation}

\noindent
where, $\mathbf{\hat{D}}=[\hat{D}_1,\hat{D}_2,\dots,\hat{D}_N]$ is the vector containing the 
attributed values $\hat{D}_r$ to each rule $r$, $\mathbf{\Psi}(\hat{u})=[\psi_1(\hat{u}),
\psi_2(\hat{u}),\dots,\psi_N(\hat{u})]$ is a vector with components $\psi_r(\hat{u})= w_r/
\sum_{r=1}^{N}w_r$ and $w_r$ is the firing strength of each rule.

In order to ensure the best possible estimate $\hat{d}(\hat{u})$, the vector of adjustable 
parameters can be automatically updated by the following adaptation law:

\begin{equation}
\mathbf{\dot{\hat{D}}}=-\gamma s\mathbf{\Psi}(\hat{u})
\label{eq:adapta}
\end{equation}

\noindent
where $\gamma$ is a strictly positive constant related to the adaptation rate. 

It is important to emphasize that the chosen adaptation law, Eq.~(\ref{eq:adapta}), must not only 
provide a good approximation to $d(u)$ but also not compromise the attractiveness of the sliding
surface, as will be proven in the following theorem.

\begin{theorem}
\label{th:theo}
Consider the uncertain nonlinear system (\ref{eq:system}) subject to the dead-zone (\ref{eq:dzone2}) 
and Assumptions \ref{as:output}--\ref{as:traj}. Then, the controller defined by (\ref{eq:usgn}), 
(\ref{eq:gain}), (\ref{eq:dcvector}) and (\ref{eq:adapta}) ensures the convergence of the tracking 
error vector to the sliding surface $S$.
\end{theorem}

\textbf{Proof:}
Let a positive-definite function $V$ be defined as

\begin{displaymath}
\displaystyle
V(t)=\frac{1}{2}s^2+\frac{bm}{2\gamma}\mathbf{\Delta^\mathrm{T}\Delta}
\end{displaymath}

\noindent
where $\mathbf{\Delta}=\mathbf{\hat{D}}-\mathbf{\hat{D}}^*$ and $\mathbf{\hat{D}}^*$ is the optimal
parameter vector, associated to the optimal estimate $\hat{d}^*(\hat{u})$. Thus, the time derivative 
of $V$ is

\begin{align*}
\dot{V}(t)&=s\dot{s}+bm\gamma^{-1}\mathbf{\Delta^\mathrm{T}\dot{\Delta}}\\
&=(\tilde{x}^{(n)}+\mathbf{\bar{c}^\mathrm{T}\tilde{x}})s
+bm\gamma^{-1}\mathbf{\Delta^\mathrm{T}\dot{\Delta}}\\
&=(x^{(n)}-x^{(n)}_d+\mathbf{\bar{c}^\mathrm{T}\tilde{x}})s
+bm\gamma^{-1}\mathbf{\Delta^\mathrm{T}\dot{\Delta}}\\
&=(f+bmu-bmd-x^{(n)}_d+\mathbf{\bar{c}^\mathrm{T}\tilde{x}})s
+bm\gamma^{-1}\mathbf{\Delta^\mathrm{T}\dot{\Delta}}\\
&=[f+bm\widehat{bm}^{-1}(-\hat{f}+x^{(n)}_d-\mathbf{\bar{c}^\mathrm{T}\tilde{x}})+bm\hat{d}
-bmK\sgn(s)-bmd-(x^{(n)}_d-\mathbf{\bar{c}^\mathrm{T}\tilde{x}})]s
+bm\gamma^{-1}\mathbf{\Delta^\mathrm{T}\dot{\Delta}}
\end{align*}

Defining the minimum approximation error as $\varepsilon=\hat{d}^*-d$, recalling that $\hat{u}=
\widehat{bm}^{-1}(-\hat{f}+x^{(n)}_d-\mathbf{\bar{c}^\mathrm{T}\tilde{x}})$, and noting that 
$\mathbf{\dot{\Delta}}=\mathbf{\dot{\hat{D}}}$ and $f=\hat{f}-(\hat{f}-f)$, $\dot{V}$ becomes:

\begin{align*}
\dot{V}(t)&=-[(\hat{f}-f)-bm\varepsilon-bm(\hat{d}-\hat{d}^*)+\widehat{bm}\hat{u}
-bm\hat{u}+bmK\sgn(s)]s+bm\gamma^{-1}\mathbf{\Delta^\mathrm{T}\dot{\hat{D}}}\\
&=-[(\hat{f}-f)-bm\varepsilon-bm(\hat{D}-\hat{D}^*)\mathbf{\Psi}(\hat{u})
+\widehat{bm}\hat{u}-bm\hat{u}+bmK\sgn(s)]s
+bm\gamma^{-1}\mathbf{\Delta^\mathrm{T}\dot{\hat{D}}}\\
&=-[(\hat{f}-f)-bm\varepsilon+\widehat{bm}\hat{u}-bm\hat{u}
+bmK\sgn(s)]s+bm\gamma^{-1}\mathbf{\Delta}^{\mathrm{T}}[\mathbf{\dot{\hat{D}}}
+\gamma s\mathbf{\Psi}(\hat{u})]
\end{align*}

By applying the adaptation law (\ref{eq:adapta}) to $\mathbf{\dot{\hat{D}}}$:

\begin{displaymath}
\dot{V}(t)=-[(\hat{f}-f)-bm\varepsilon+\widehat{bm}^{-1}\hat{u}-bm\hat{u}+bmK\sgn(s)]s
\end{displaymath}

Furthermore, considering Assumptions~\ref{as:dband}--\ref{as:bbounds}, defining $K$ according to 
(\ref{eq:gain}) and verifying that $|\varepsilon|=|\hat{d}^*-d|\le|\hat{d}-d|\le|\hat{d}|
+\delta$, it follows that 

\begin{equation}
\dot{V}(t)\le-\eta|s|
\label{eq:dv}
\end{equation}

\noindent
which implies $V(t)\le V(0)$ and that $s$ and $\mathbf{\Delta}$ are bounded. Considering that 
$s(\mathbf{\tilde{x}})=\mathbf{c^\mathrm{T}\tilde{x}}$, it can be verified that $\mathbf{\tilde{x}}$
is also bounded. Hence, Eq.~(\ref{eq:sd}) and Assumption~\ref{as:traj} implies that $\dot{s}$ is also 
bounded.

Integrating both sides of (\ref{eq:dv}) shows that

\begin{displaymath}
\displaystyle
\lim_{t\rightarrow\infty}\int_0^t\eta|s|\,d\tau \le \lim_{t\rightarrow\infty}\left[V(0)-V(t)\right]
\le V(0) < \infty
\end{displaymath}

Since the absolute value function is uniformly continuous, it follows from Barbalat's lemma 
\cite{khalil1} that $s\to0$ as $t\to\infty$, which ensures the convergence of the tracking 
error vector to the sliding surface $S$ and completes the proof.\hfill$\square$ \vspace*{12pt}

However, the presence of a discontinuous term in the control law leads to the well known chattering 
phenomenon. To overcome the undesirable chattering effects, \citep{slotine2} proposed the 
adoption of a thin boundary layer, $S_\phi$, in the neighborhood of the switching surface:

\begin{displaymath}
\displaystyle
S_\phi=\big\{\mathbf{\tilde{x}}\in\mathbb{R}^n\:\big|\:|s(\mathbf{\tilde{x}})|\le\phi\big\}
\end{displaymath}

\noindent
where $\phi$ is a strictly positive constant that represents the boundary layer thickness.

The boundary layer is achieved by replacing the sign function by a continuous interpolation inside 
$S_\phi$. It should be noted that this smooth approximation, which will be called here $\varphi(s,
\phi)$, must behave exactly like the sign function outside the boundary layer. There are several 
options to smooth out the ideal relay but the most common choices are the saturation function:

\begin{displaymath}
\displaystyle
\sat(s/\phi) = \left\{\begin{array}{cc}
\sgn(s)&\mbox{if}\quad |s/\phi|\ge1 \\
s/\phi&\mbox{if}\quad |s/\phi|<1 
\end{array}\right.
\end{displaymath}

\noindent
and the hyperbolic tangent function $\tanh(s/\phi)$.

In this way, to avoid chattering, a smooth version of Eq.~(\ref{eq:usgn}) can be adopted:

\begin{equation}
u=\widehat{bm}^{-1}(-\hat{f}+x^{(n)}_d-\mathbf{\bar{c}^\mathrm{T}\tilde{x}})
+\hat{d}(\hat{u})-K\varphi(s,\phi)
\label{eq:usmooth}
\end{equation}

Nevertheless, it should be emphasized that the substitution of the discontinuous term by a smooth 
approximation inside the boundary layer turns the perfect tracking into a tracking with guaranteed 
precision problem, which actually means that a steady-state error will always remain. 

\begin{remark}
It has been demonstrated by \citep{ijac2009} that by adopting a smooth sliding mode controller, 
the tracking error vector will exponentially converge to a closed region $\Phi=\{\mathbf{\tilde{x}}
\in\mathbb{R}^n\:|\:|s(\mathbf{\tilde{x}})|\le\phi\mbox{ and }|\tilde{x}^{(i)}|\le\zeta_i\lambda^{i
-n+1}\phi,i=0,1,\ldots,n-1\}$, with $\zeta_i$ defined as
\begin{displaymath}
\displaystyle
\zeta_i = \left\{\begin{array}{cl}
1&\mbox{for}\quad i=0 \\
1+\sum^{i-1}_{j=0}\binom{i}{j}\zeta_j&\mbox{for}\quad i=1,2,\ldots,n-1.
\end{array}\right.
\end{displaymath}
\end{remark}

\section{ILLUSTRATIVE EXAMPLE: ELECTRO-HYDRAULIC SYSTEM}

Electro-hydraulic actuators play an essential role in several branches of industrial activity and are 
frequently the most suitable choice for systems that require large forces at high speeds. Their 
application scope ranges from robotic manipulators to aerospace systems. Another great advantage of 
hydraulic systems is the ability to keep up the load capacity, which in the case of electric actuators 
is limited due to excessive heat generation. 

\begin{figure}[htb]
\centering
\includegraphics[width=0.8\textwidth]{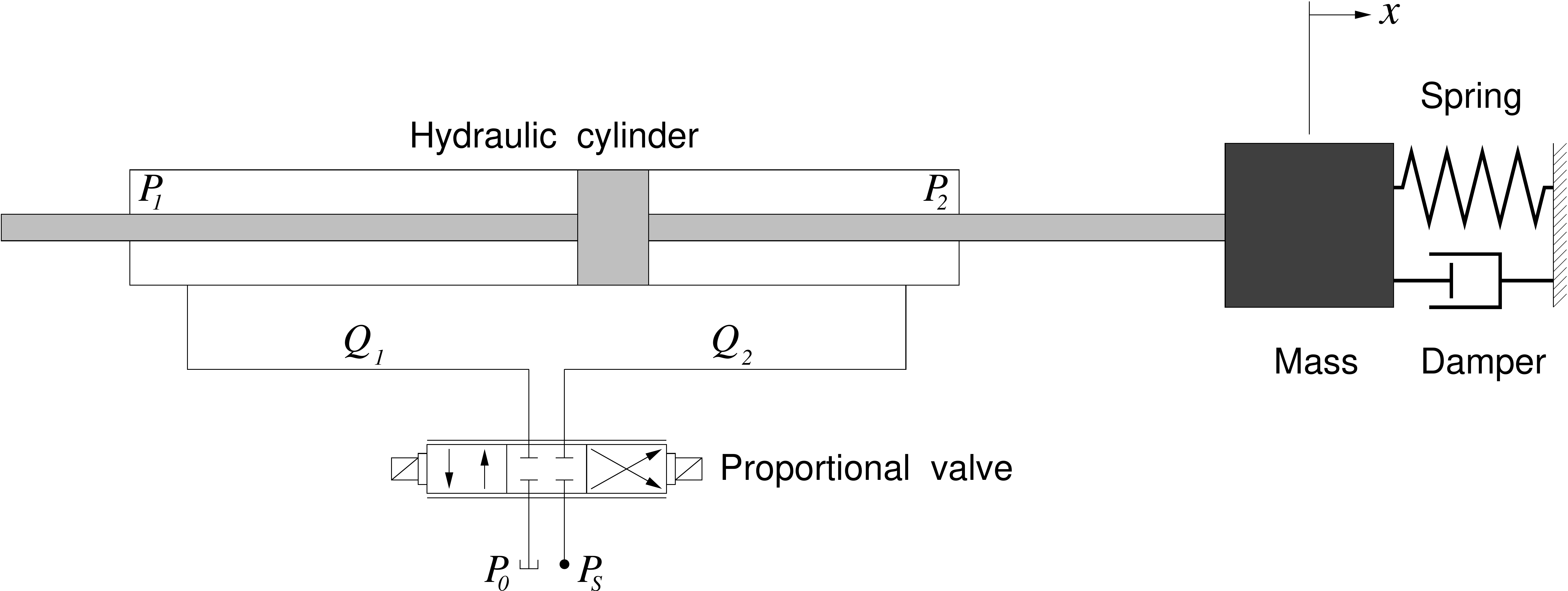}
\caption{Schematic diagram of the electro-hydraulic servo-system.}
\label{fig:sistema}
\end{figure}

The electro-hydraulic system considered in this work consists of a four-way proportional valve, 
a hydraulic cylinder and variable load force. The variable load force is represented by a 
mass--spring--damper system. The schematic diagram of the system under study is presented in
Fig.~\ref{fig:sistema}.

The dynamic behavior of electro-hydraulic systems is highly nonlinear, which in fact makes the design 
of controllers for such systems a challenge for the conventional and well established linear control 
methodologies. In addition to the common nonlinearities that originate from the compressibility 
of the hydraulic fluid and valve flow-pressure properties, most electro-hydraulic systems are also 
subjected to hard nonlinearities such as dead-zone due to valve spool overlap. Considering the voltage 
as control input $u$ and the valve gain as dead-zone slope $m$, the valve nonlinearity can be 
mathematically described by Eqs.~(\ref{eq:dzone2})--(\ref{eq:dsat}), with parameters $\delta_l$ and 
$\delta_r$ depending on the size of the overlap region.

In this way, the mathematical model that represents the electro-hydraulic system can be stated as
follows \cite{diname2009}:

\begin{displaymath}
\dddot{x}=-\mathbf{a^\mathrm{T}x}+b(\mathbf{x},u)u-b(\mathbf{x},u)d(u)
\end{displaymath}

\noindent
where $\mathbf{x}=[x,\dot{x},\ddot{x}]$ is the state vector with an associated coefficient vector 
$\mathbf{a}=[a_0,a_1,a_2]$ defined according to

\begin{displaymath}
\displaystyle
a_0=\frac{4\beta_eC_{tp}K_s}{V_tM_t}\quad;\quad
a_1=\frac{K_s}{M_t}+\frac{4\beta_eA_{p}^2}{V_tM_t}+\frac{4\beta_eC_{tp}B_t}{V_tM_t}\quad;\quad
a_2=\frac{B_t}{M_t}+\frac{4\beta_eC_{tp}}{V_t}
\end{displaymath}

\noindent
and

\begin{displaymath}
\displaystyle
b(\mathbf{x},u)=\frac{4\beta_eA_p}{V_tM_t}C_dwm\sqrt{\frac{1}{\rho}\big[P_s-\sgn(u)
\big(M_t\ddot{x}+B_t\dot{x}+K_s{x}\big)/A_p\big]}
\end{displaymath}

Here, $x$ is the piston displacement, $M_t$ the total mass of piston and load referred to piston, 
$B_t$ the viscous damping coefficient of piston and load, $K_s$ the load spring constant, $A_p$ 
the ram area of the two chambers (symmetrical cylinder), $C_{tp}$ the total leakage coefficient of 
piston, $V_t$ the total volume under compression in both chambers, $\beta_e$ the effective bulk 
modulus, $C_d$ the discharge coefficient, $w$ the valve orifice area gradient, $\rho$ the hydraulic 
fluid density and $P_s$ the supply pressure 

On this basis, according to the previously described control scheme and considering $s=\ddot{\tilde{x}}
+2\lambda\dot{\tilde{x}}+\lambda^2\tilde{x}$, the smooth control law can be defined as follows

\begin{displaymath}
u=\hat{b}^{-1}(\mathbf{\hat{a}^\mathrm{T}x}+\dddot{x}_d-2\lambda\ddot{\tilde{x}}-\lambda^2
\dot{\tilde{x}})+\hat{d}(\hat{u})-K\sat(s/\phi) 
\end{displaymath}

The simulation studies were performed with a numerical implementation in C, with sampling rates of 
400 Hz for control system and 800 Hz for dynamic model. The adopted parameters for the electro-hydraulic 
system were $P_s=7$ MPa, $\rho=850$ kg/m$^3$, $C_d=0.6$, $w=2.5\times10^{-2}$ m, $A_p=3\times10^{-4}$ 
m$^2$, $C_{tp}=2\times10^{-12}$ m$^3$/(s Pa), $\beta_e=700$ MPa, $V_t=6\times10^{-5}$ m$^3$, $M_t=250$ 
kg, $B_t=100$ Ns/m, $K_s=75$ N/m, $k_l=1.8\times10^{-6}$ m/V, $k_r=2.2\times10^{-6}$ m/V, $\delta_l=
-1.1$ V and $\delta_r=0.9$ V. For controller parameters, the following values were chosen $\lambda=8$, 
$\varphi=4$, $\gamma=1.2$, $\delta=1.1$, $\phi=1$, $\eta=0.1$ and $\alpha=0$. 

\begin{figure}[htb]
\centering
\includegraphics[width=0.6\textwidth]{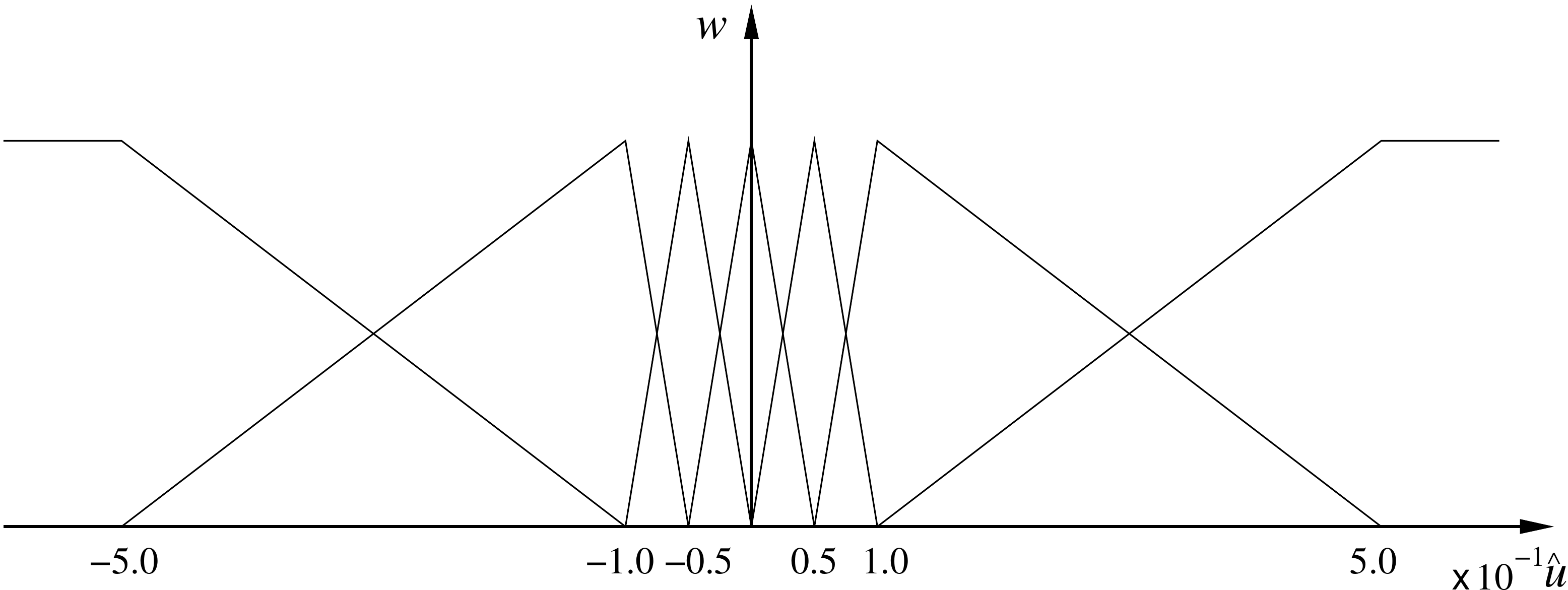}
\caption{Adopted fuzzy membership functions.} 
\label{fi:fset}
\end{figure}

Concerning the fuzzy inference system, triangular and trapezoidal membership functions, respectively 
$\mu_\mathrm{tri}$ and $\mu_\mathrm{trap}$, were adopted for $\hat{U}_r$, with central values defined
as $C=\{-5.0\:;\:-1.0\:;\:-0.5\:;\:0.0\:;\:0.5\:;\:1.0\:;\:5.0\}\times10^{-1}$ (see Fig.~\ref{fi:fset}).
It is also important to emphasize, that the vector of adjustable parameters was initialized with zero 
values, $\mathbf{\hat{D}=0}$, and updated at each iteration step according to the adaptation law 
presented in Eq.~(\ref{eq:adapta}). 

\begin{figure}[htb]
\centering
\mbox{
\subfigure[Tracking performance.]{\label{fi:graf1_1} 
\includegraphics[width=0.35\textwidth]{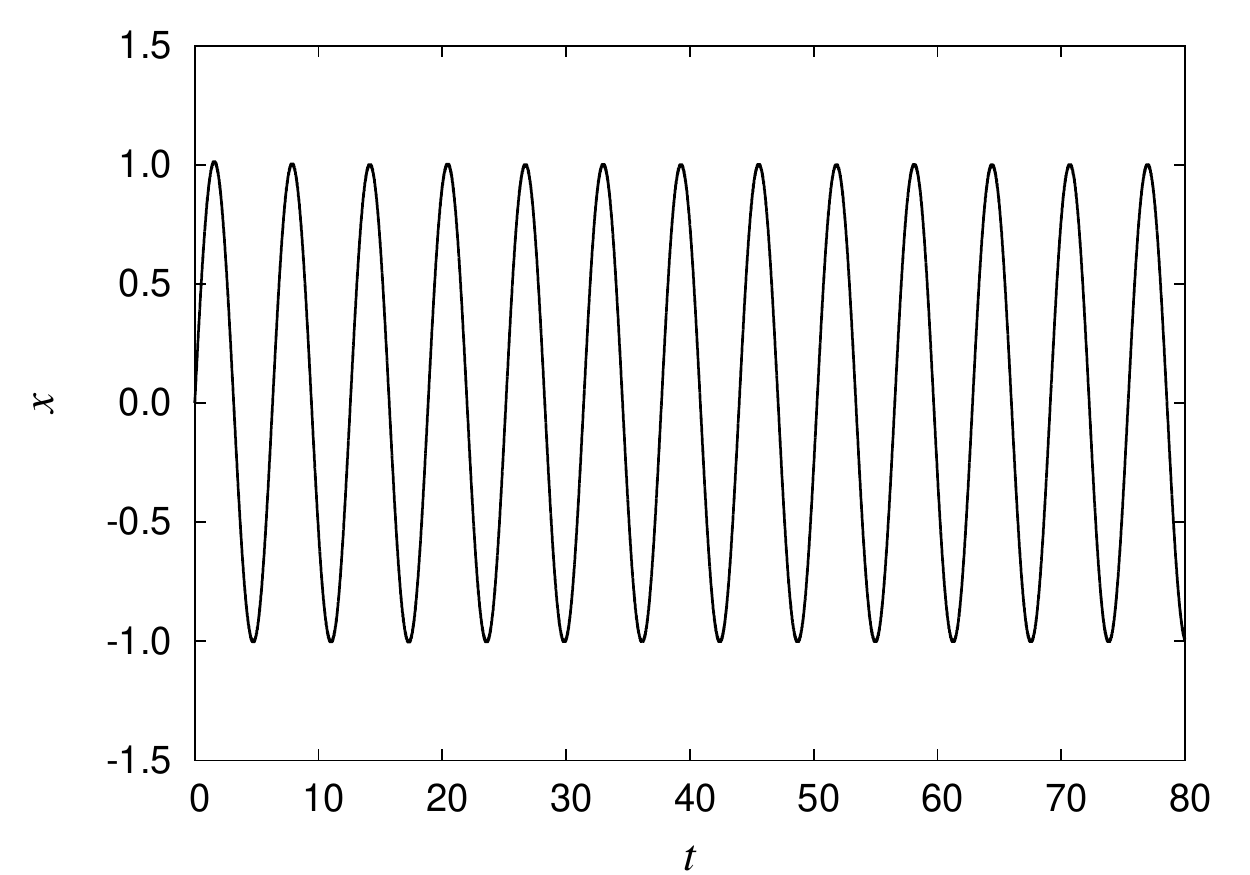}}
\subfigure[Control voltage.]{\label{fi:graf2_1} 
\includegraphics[width=0.35\textwidth]{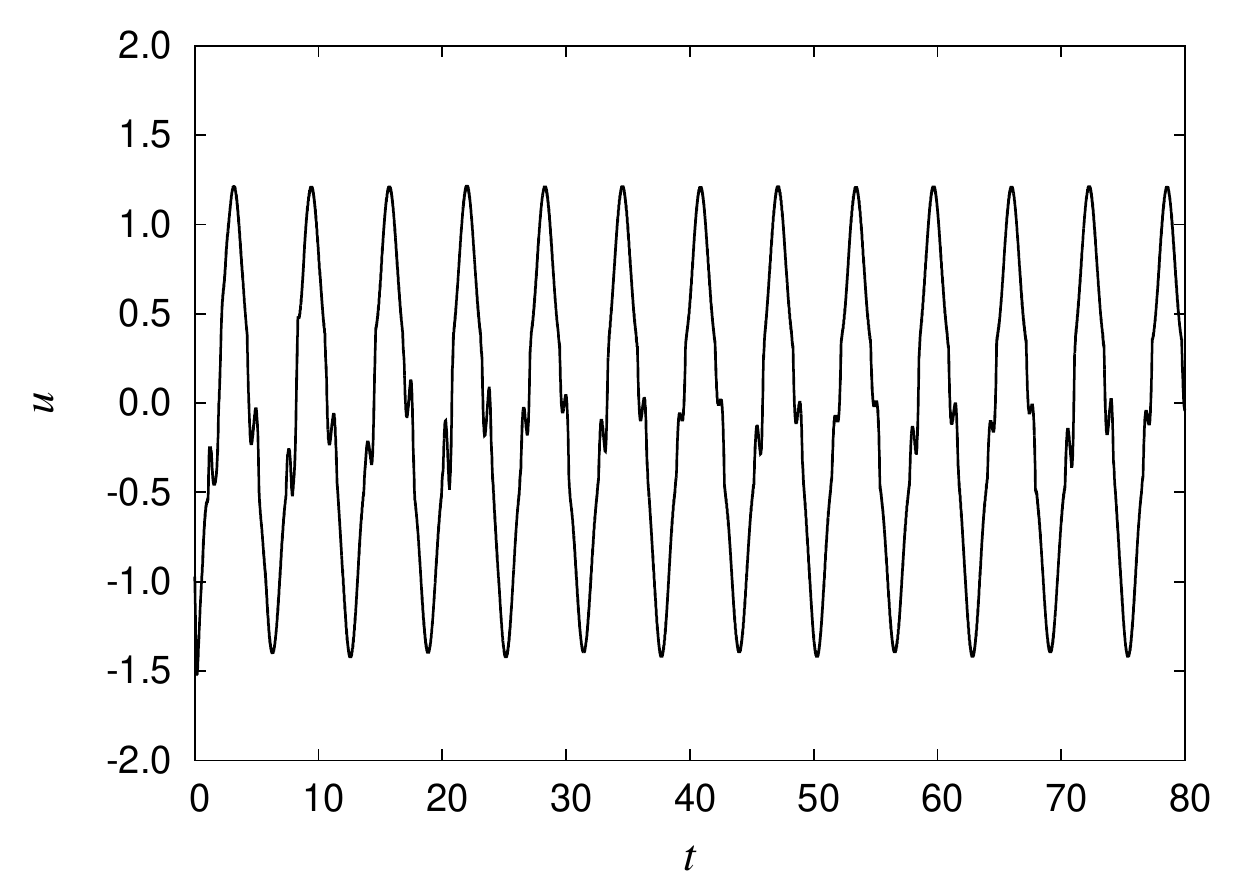}}
}
\mbox{
\subfigure[Tracking error.]{\label{fi:graf3_1} 
\includegraphics[width=0.35\textwidth]{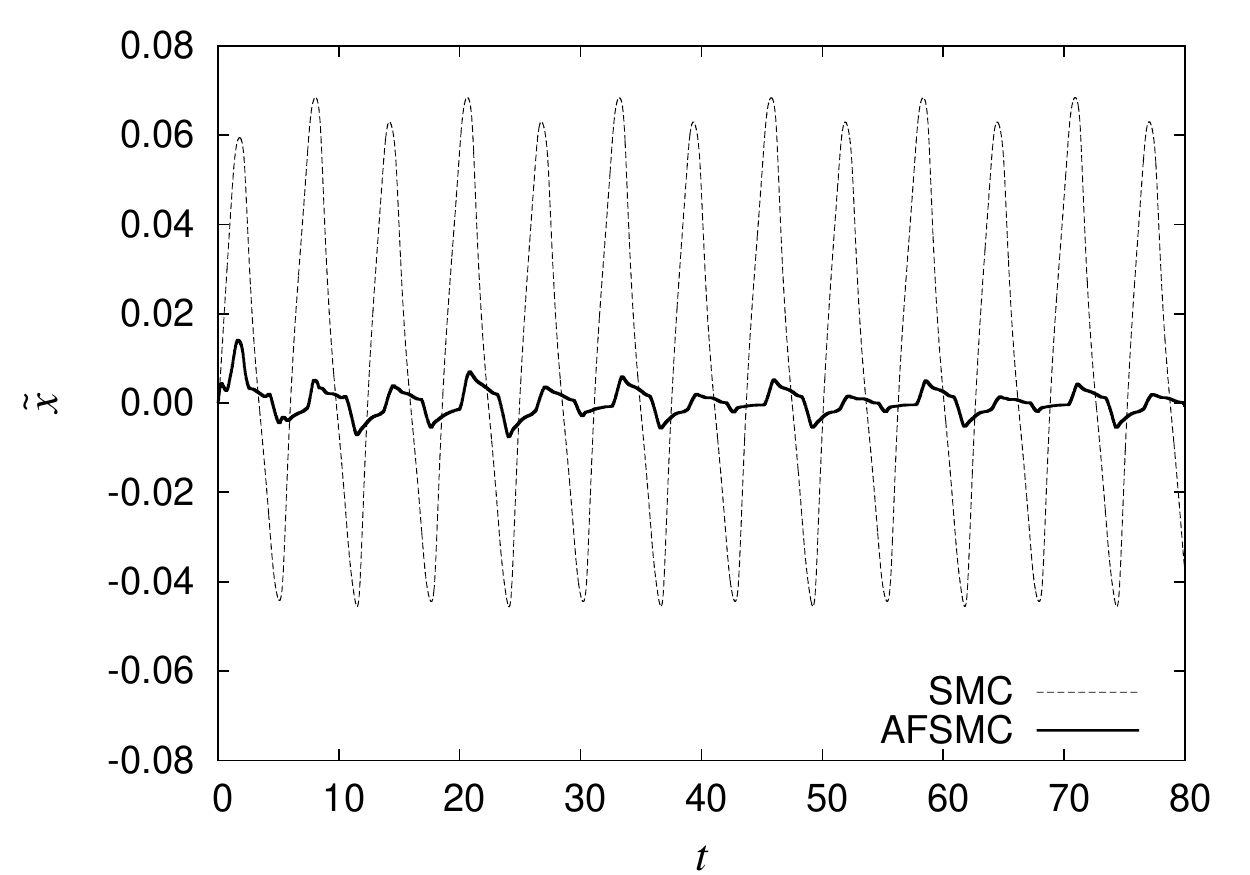}}
\subfigure[Convergence of $\hat{d}(\hat{u})$ to $d(u)$.]{\label{fi:graf4_1} 
\includegraphics[width=0.35\textwidth]{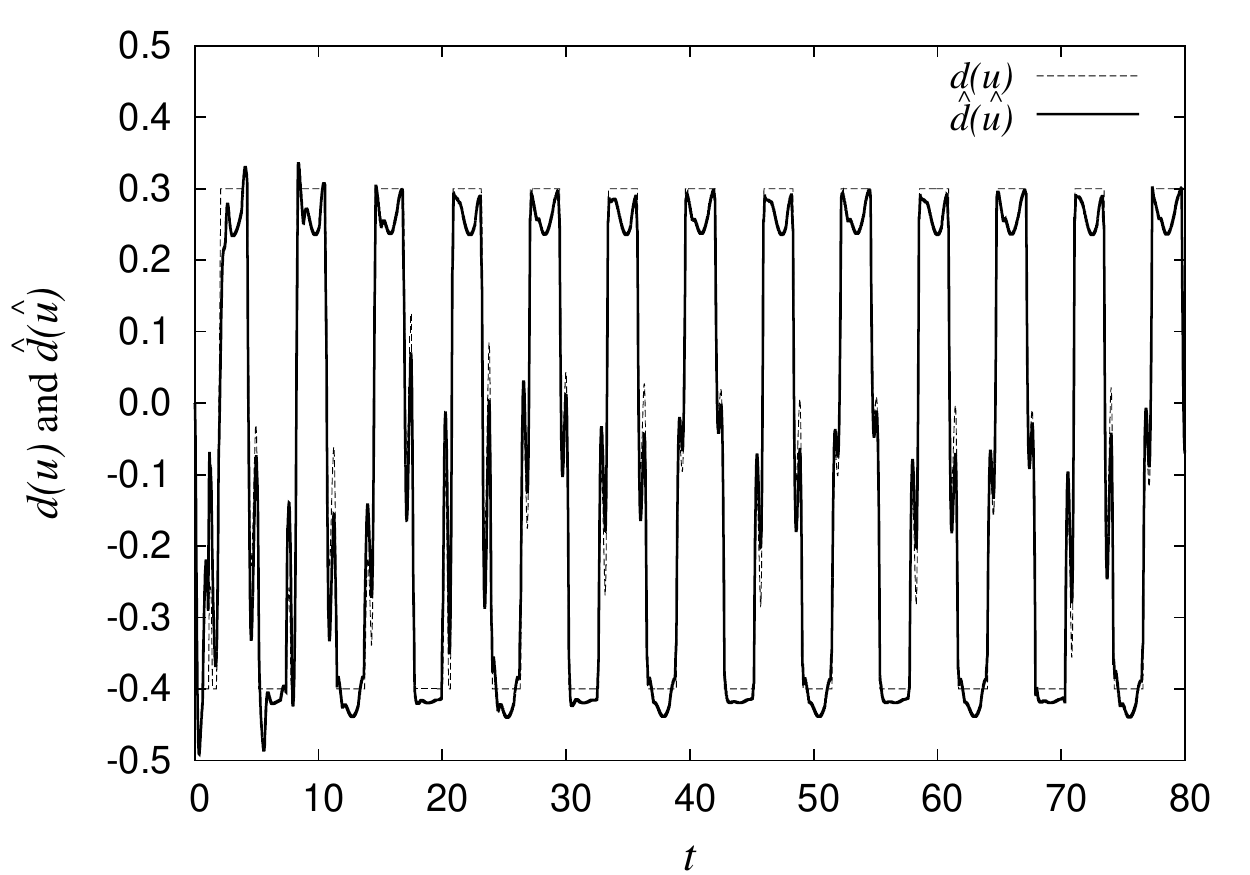}}
}
\caption{Tracking performance with unknown dead-zone and well known model parameters.}
\label{fi:sim1}
\end{figure}

In order to evaluate the control system performance, two numerical simulations were carried out. 
In the first case, it was assumed that the model parameters were perfectly known but the dead-zone 
width was considered unknown. Figure~\ref{fi:sim1} shows the obtained results for the tracking of
$x_d=0.5\sin(0.1t)$ m. 

As observed in Fig.~\ref{fi:sim1}, the adaptive fuzzy sliding mode controller (AFSMC) is able to
provide trajectory tracking with small associated error and no chattering at all. It can be also 
verified that the proposed control law leads to a smaller tracking error when compared with the 
conventional sliding mode controller (SMC), Fig.~\ref{fi:graf3_1}. The improved performance of 
AFSMC over SMC is due to its ability to compensate for dead-zone effects, Fig.~\ref{fi:graf4_1}.
The AFSMC can be easily converted to the classical SMC by setting the adaptation rate to zero, 
$\varphi=0$. 

\begin{figure}[htb]
\centering
\mbox{
\subfigure[Tracking performance.]{\label{fi:graf1_2} 
\includegraphics[width=0.35\textwidth]{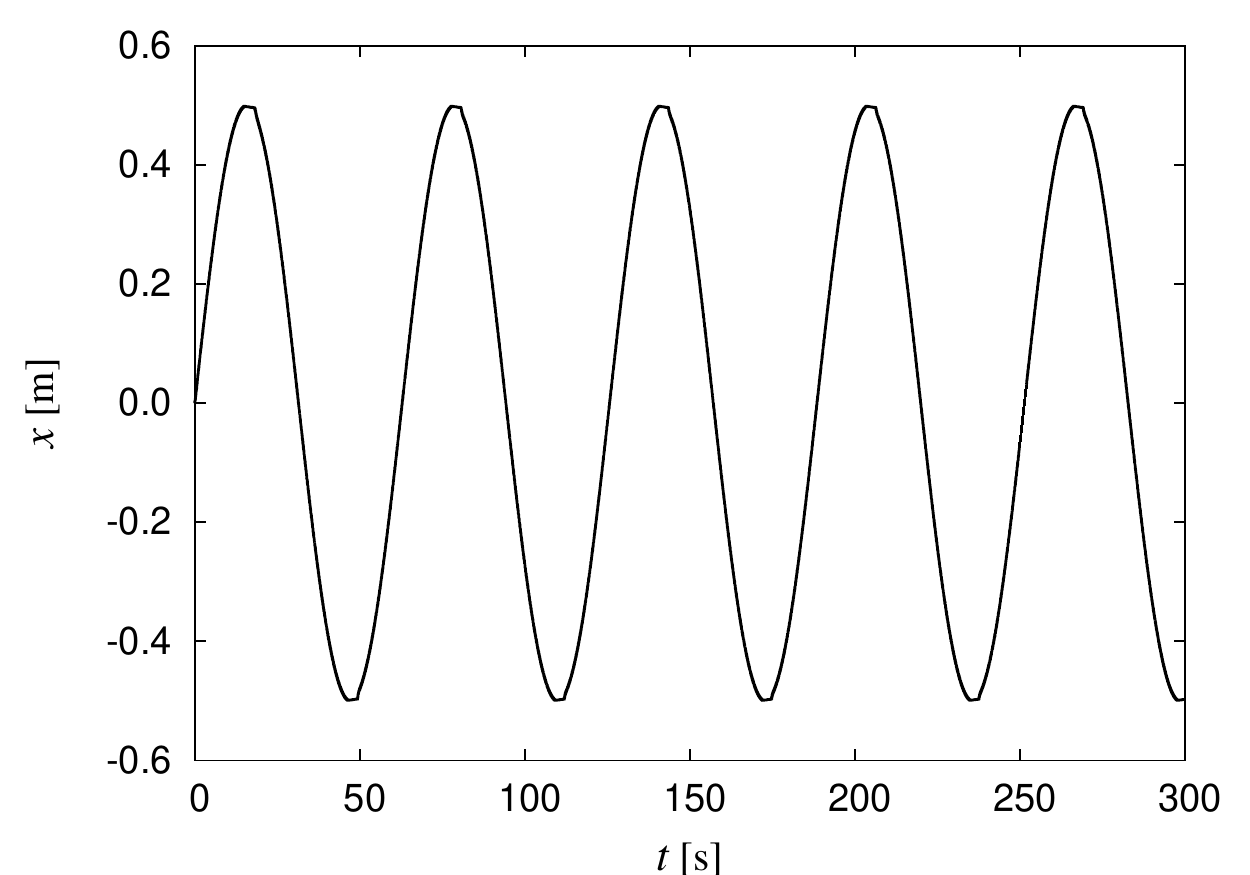}}
\subfigure[Control voltage.]{\label{fi:graf2_2} 
\includegraphics[width=0.35\textwidth]{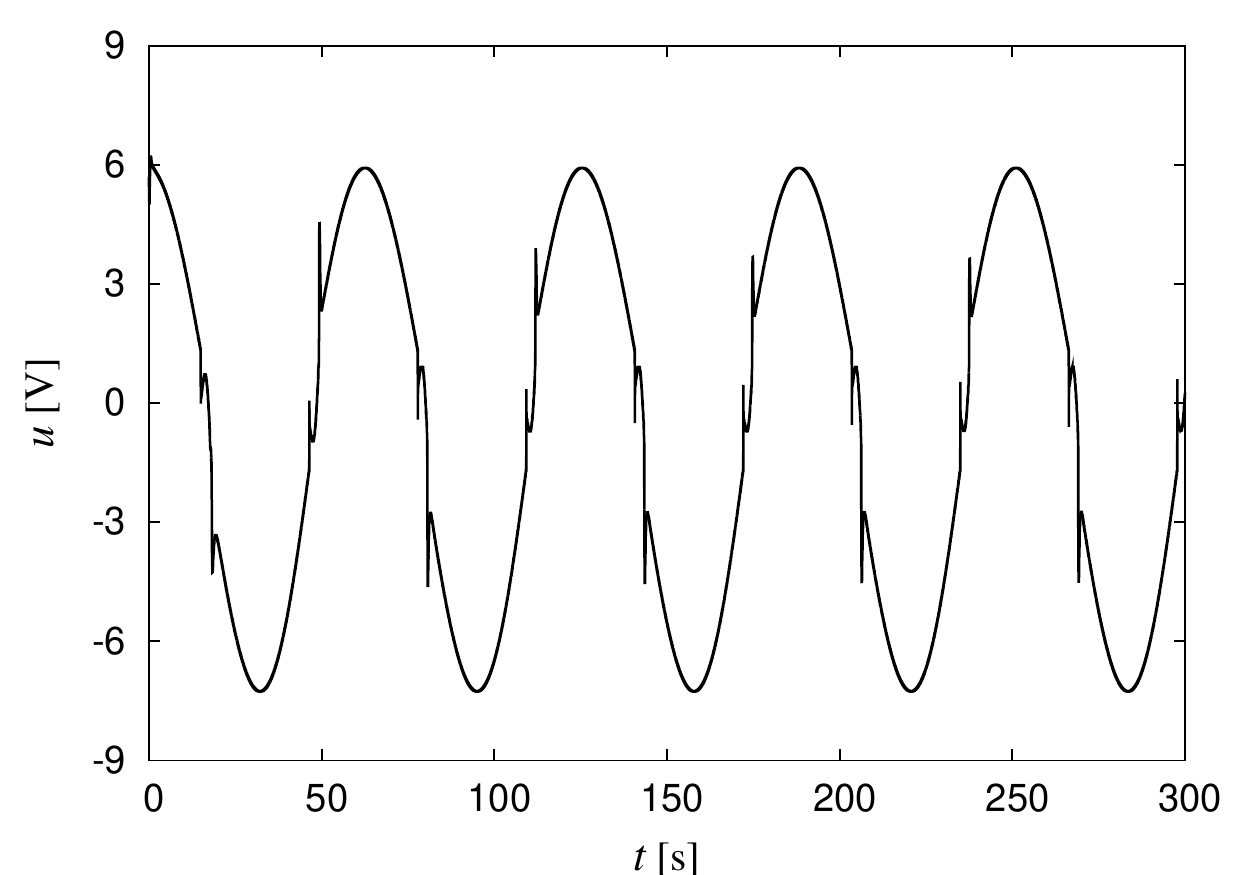}}
}
\mbox{
\subfigure[Tracking error.]{\label{fi:graf3_2} 
\includegraphics[width=0.35\textwidth]{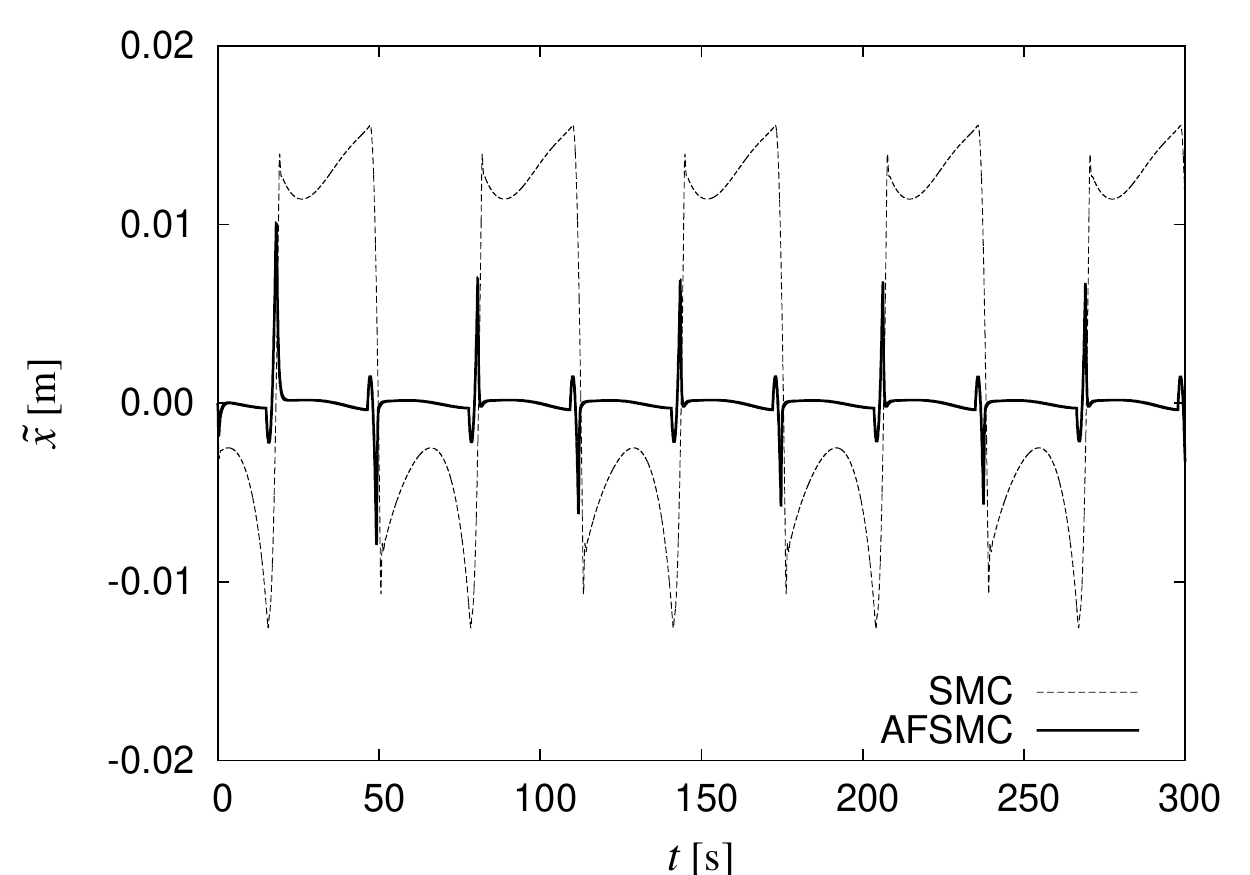}}
\subfigure[Convergence of $\hat{d}(\hat{u})$ to $d(u)$.]{\label{fi:graf4_2} 
\includegraphics[width=0.35\textwidth]{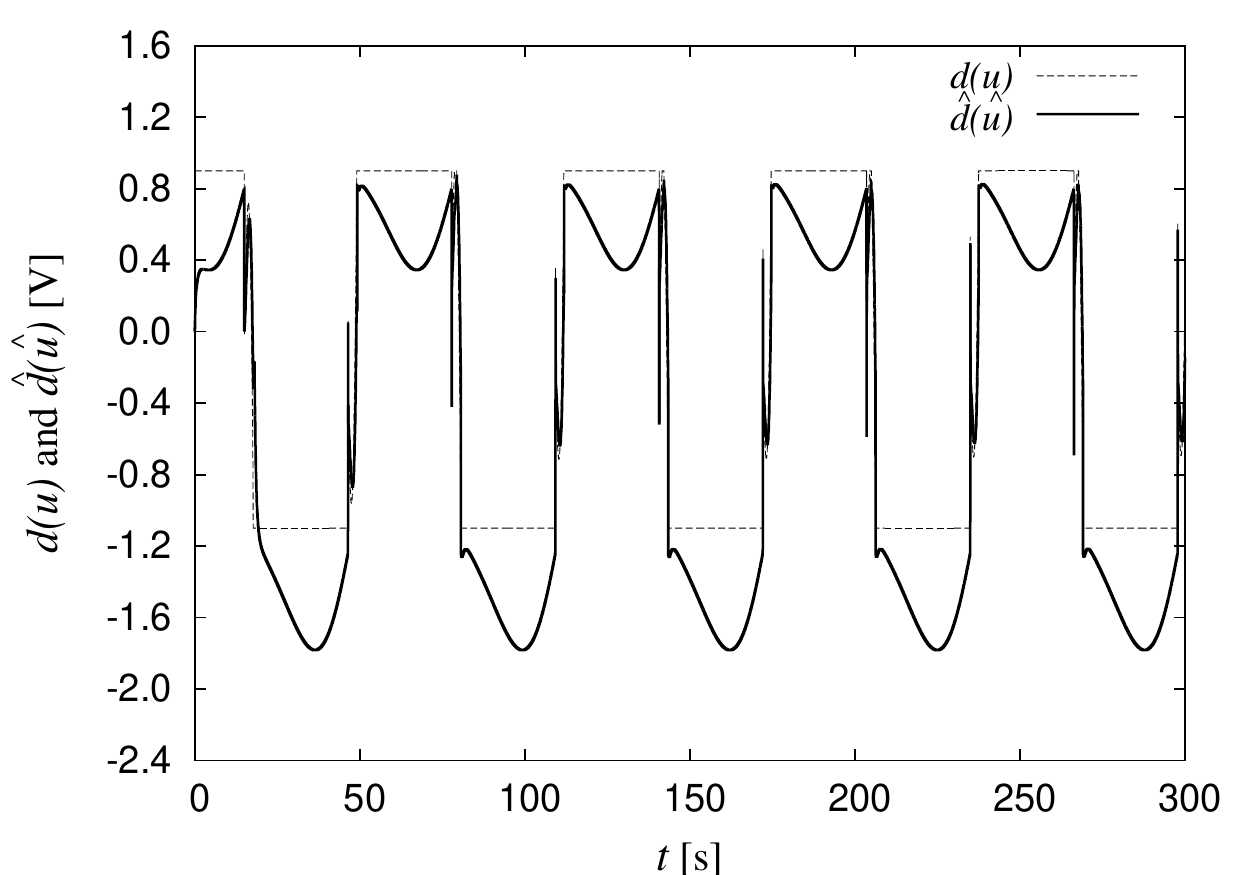}}
}
\caption{Tracking performance with unknown dead-zone and uncertain model parameters.}
\label{fi:sim2}
\end{figure}

In the second simulation study it was assumed that the model parameters are not exactly known. On 
this basis, considering a maximal uncertainty of $\pm10\%$ over the value of $k_v$ and variations 
of $\pm20\%$ in the supply pressure, $P_s=7(1+0.2\sin(x))$ MPa, the estimates $\hat{k}_v=2\times
10^{-6}$ m/V and $\hat{P}_s=7$ MPa were chosen for the computation of $\hat{b}$ in the control law. 
The other model and controller parameters, as well as the desired trajectory, were chosen as before. 
The obtained results are presented in Fig.~\ref{fi:sim2}.

Despite the dead-zone input and uncertainties with respect to model parameters, the AFSMC allows 
the electro-hydraulic actuated system to track the desired trajectory with a small tracking error, 
(see Fig.~\ref{fi:sim2}). As before, the undesirable chattering effect is not observed, 
Fig.~\ref{fi:graf2_2}. Through the comparative analysis shown in Fig.~\ref{fi:graf3_2}, the 
improved performance of the AFSMC over the uncompensated counterpart can be also clearly ascertained. 

\section{CONCLUSIONS}

The present work addresses the problem of controlling uncertain nonlinear systems subject to a
non-symmetric dead-zone input. An adaptive fuzzy sliding mode controller is proposed to deal with 
the trajectory tracking problem. The convergence properties of the closed-loop system are 
analytically proven using Lyapunov stability theory and Barbalat's lemma. To illustrate the 
controller design method and to evaluate its performance, the proposed scheme is applied to 
an electro-hydraulic system. Through numerical simulations, the improved performance over the 
conventional sliding mode controller is also demonstrated.

\section{ACKNOWLEDGEMENTS}

The author acknowledges the support of the Brazilian Research Council (CNPq).

\end{document}